\documentstyle[11pt,newpasp,twoside,epsf]{article} 
\input psfig

\def\edcomment#1{\iffalse\marginpar{\raggedright\sl#1\/}\else\relax\fi}
\reversemarginpar

\begin{document}
\title{HI gas disks in elliptical galaxies }
\author{Elaine M. Sadler }
\affil{School of Physics, University of Sydney, NSW 2006, Australia }
\author{Tom Oosterloo and Raffaella Morganti }
\affil{NFRA, Postbus 2, 9700AA Dwingeloo, The Netherlands }

\begin{abstract}
We discuss a class of low--luminosity E/S0 galaxies which have both HI disks
and (in contrast to more luminous E/S0s with HI) ongoing star formation. 
We suggest that such objects are common, but that only a few are known at 
present because optical magnitude--limited galaxy catalogues are biased against
them. The HI Parkes All--Sky Survey (HIPASS) should eventually detect many
more. We suggest that `boxy' and `disky' ellipticals are distinct not only 
in their structure and kinematics, but in their star--formation history. 
\end{abstract}

\vspace*{-0.3cm}
\section{Introduction}
Twenty years ago, it might have been surprising to see elliptical galaxies
featured at a conference on ``Galaxy Disks and Disk Galaxies''. Now, however,
it is generally accepted that many elliptical galaxies have stellar and/or
gaseous disks.  Intermediate--luminosity ellipticals often show `disky'
isophotes and/or rapid stellar rotation suggesting the presence of an inner
stellar disk (e.g. Bender et al.\ 1988; Rix \& White 1990). Extended disks of
gas and dust are often seen around giant ellipticals (e.g. Knapp et al.\ 
1985; Sadler \& Gerhard 1986), and are usually ascribed to a recent accretion 
event or merger. Small central disks of ionized gas are also common in 
elliptical and S0 galaxies (e.g. Phillips et al.\ 1986; Buson et al.\ 1993). 

The existence of these disks has implications for our ideas about how
early--type galaxies form and evolve.  There is no clear photometric 
boundary between the E and S0 classes (e.g. van den Bergh 1989), and it has  
been suggested that `disky' ellipticals form a continuous photometric sequence 
with S0s and spiral bulges and may have a different formation mechanism from 
the (generally more massive) `boxy' ellipticals (Capaccioli et al.\ 1990; 
Kormendy \& Bender 1996). 

For the past few years, we (and others) have been working to 
obtain high--quality HI images for a larger sample of early--type galaxies 
in order to learn more about the relationship between gas content, galaxy 
evolution, environment and the triggering of an active nucleus.  
Jacqueline van Gorkom (this meeting) has reviewed what we now know about 
the HI structure and kinematics in early--type galaxies.  In this paper, we 
focus on two topics --- star formation in the HI disks of low luminosity
ellipticals, and the likely impact of the HI Parkes All--Sky Survey (HIPASS) 
on HI studies of early--type galaxies. 
In the discussion which follows, we assume H$_\circ$=50 km s$^{-1}$ Mpc$^{-1}$, 
and that a `low--luminosity' galaxy has M$_{\rm B} > -20$\ mag. 

\vspace*{-0.24cm}
\section{HI disks and star formation in low--luminosity early--type galaxies }
Lake \& Schommer (1984) first showed that HI was common in 
low--luminosity E/S0 galaxies, detecting 11/28 such systems (39\%) in their 
HI survey at Arecibo.  Four of these galaxies were later imaged with the VLA 
and shown to have regular velocity fields (Lake et al.\ 1987). 
At about the same time, Phillips et al. (1986) found HII region--like emission 
line ratios in the spectra of 5/18 low--luminosity E/S0s (28\%), implying 
that these galaxies are currently forming massive stars.  In contrast, 
none of the $\sim$200 luminous E/S0s observed by Phillips et al.\ showed 
evidence for the presence of HII regions.  

Luminosity functions (e.g. Binggeli et al.\ 1988) show that low--luminosity 
E/S0 galaxies are more common than luminous ones in terms of their space 
density, yet they remain less well--studied because optical galaxy catalogues 
are strongly biased towards optically--luminous galaxies.  Many of the 
low--luminosity galaxies which have been studied lie in clusters (because 
they are easier to find there) and may not be typical of the field population. 

Sadler et al.\ (2000) studied the HI distribution and kinematics in four
low--luminosity field E/S0 galaxies shown by Phillips et al.\ (1986) to have
HII region--like optical emission--line spectra.  The HI typically extended to
2--3 times the optical radius (see Figure 1), and the velocity fields were
characteristic of settled disks with regular rotation. In two of the four
galaxies the HI rotation axis was misaligned with the photometric axis of the
optical galaxy.  In all four galaxies the HI was very centrally concentrated,
in contrast to early--type spiral bulges and luminous ellipticals which often
show central HI holes. 

\begin{figure}
\centerline{\psfig{figure=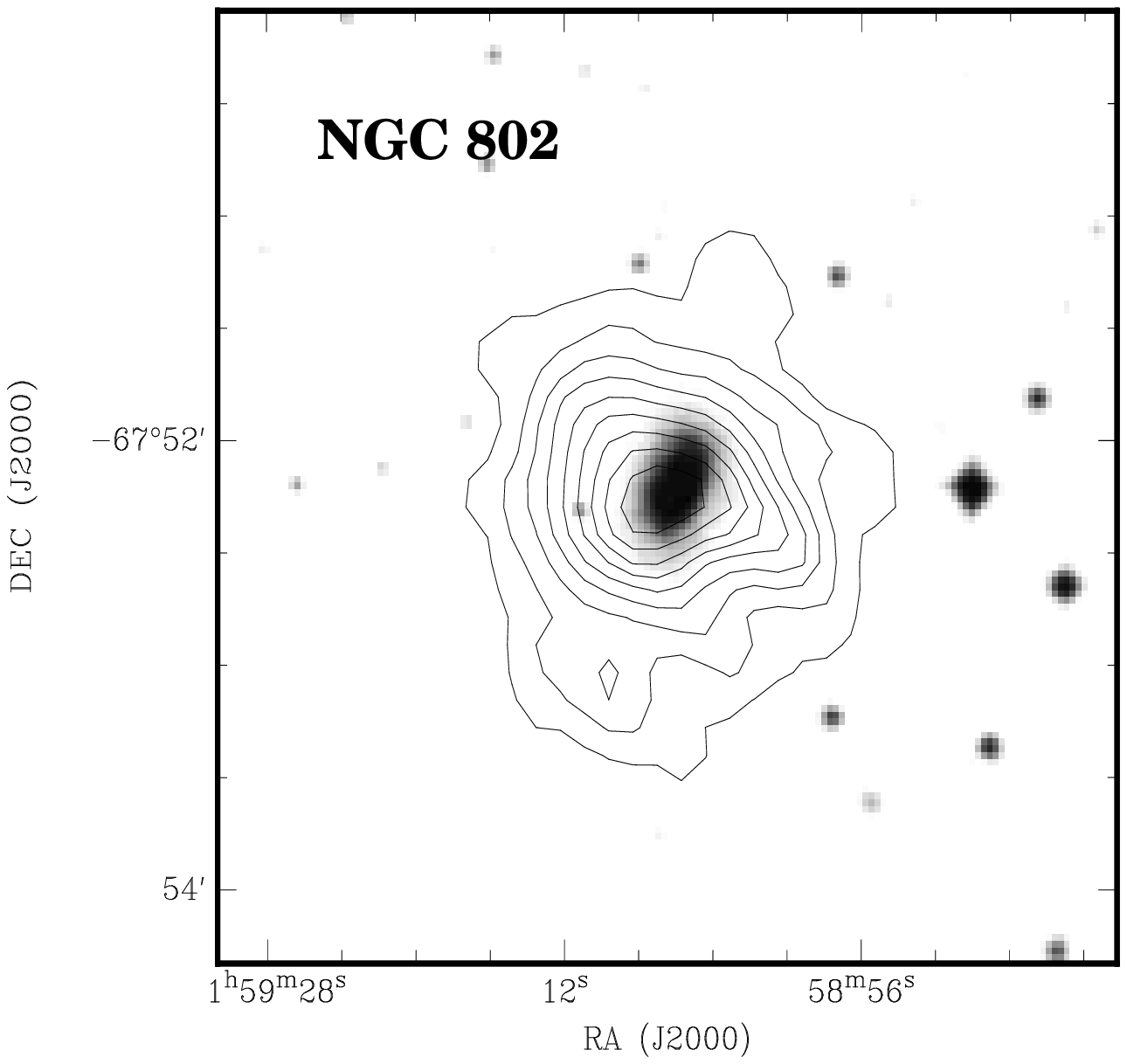,width=5cm,angle=-90}
\psfig{file=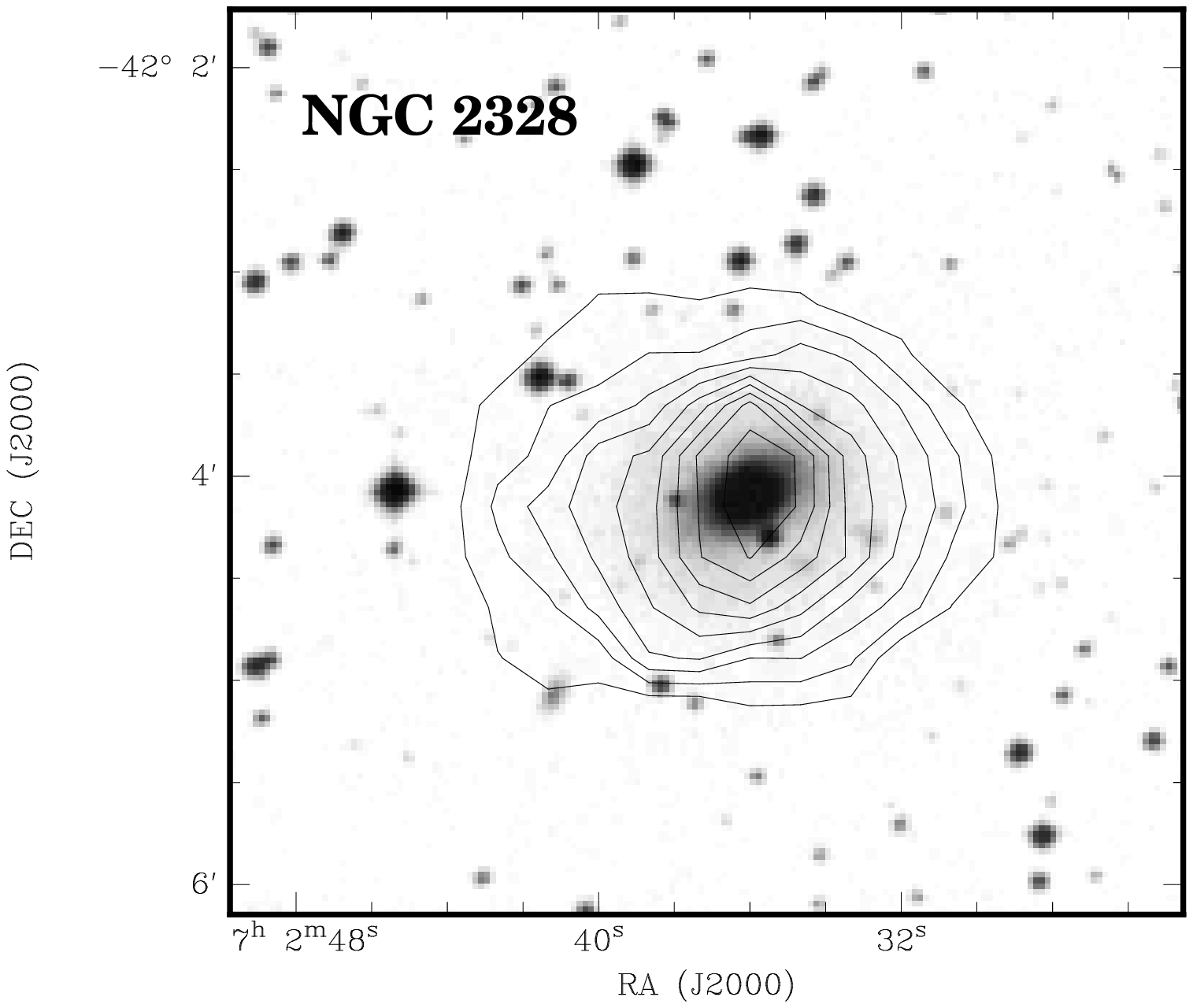,width=5cm,angle=-90}}
\caption{Total HI contours overlaid on the optical images of two 
small E/S0 galaxies: (left) NGC\,802 and (right) NGC\,2328} 
\end{figure}

Table 1 shows the current star--formation rate (SFR) for the four galaxies
studied by Sadler et al.\ (2000), derived from the H$\alpha$ emission--line
flux. Optical images and spectroscopy show that massive star formation 
in these galaxies takes place in the central 0.5--1.0\,kpc, where the HI 
surface density is highest.  Thus the bulk of star formation is concealed 
within the central bulge of these galaxies.  
The star--formation rates in Table 1 are low compared to those in 
spiral disks (e.g. around 4 M$_\odot$/yr in the Galaxy), and 
the currently--available gas supply could sustain the current rate of star 
formation for many Gyr.  The UBV colours of the galaxies in Table 1 can be 
fitted by stellar--population models with a slowly--declining star--formation 
rate S $\propto e^{-{\rm t}}$ (Searle et al.\ 1973), i.e. intermediate 
between the giant ellipticals (S $\propto e^{-10{\rm t}}$ and late--type 
spiral disks (S $\sim$ constant).  Thus all the data so far are consistent with 
a picture in which these low--luminosity E/S0 galaxies have been forming stars
at a slowly--declining rate over a Hubble time.  If so, it appears that 
bulges can be built up slowly over time, and that low--luminosity early--type 
galaxies may have a very different star--formation history from the giant 
ellipticals.  

\begin{table}[t]
\caption{Gas content, star--formation rates and gas--depletion times in four
low--luminosity E/S0 galaxies } 
\begin{tabular}{lccccl}
&&&&& \\
\tableline 
Galaxy & M$_{\rm B}$ & SFR & HI mass & M(HI) & 
HI depletion \\
       & (mag) & (M$_\odot$/yr) & (M$_\odot$) & /L$_{\rm B}$ & 
time (yr) \\
\tableline
NGC\,802       & $-18.0$ & 0.02 & 7.6$\times10^8$ & 0.42 & 
7$\times10^{10}$ \\
ESO\,027--G21  & $-19.9$ & 0.07 & 4.7$\times10^9$ & 0.40 & 
8$\times10^{10}$ \\
ESO\,118--G34  & $-17.9$ & 0.29 & 2.5$\times10^8$ & 0.12 & 
7$\times10^{8}$ \\
NGC\,2328      & $-18.7$ & 0.41 & 2.0$\times10^8$ & 0.07 & 
2$\times10^{9}$ \\
\tableline
\tableline
\end{tabular}
\end{table}

However, we are far from having a complete understanding of the evolutionary
history of low--luminosity E/S0 galaxies. It is still unclear whether the HI
commonly seen in these galaxies is mainly primordial, or whether in some cases
at least it has been accreted later on (as the misalignment of HI rotation 
axes and optical photometric axes might indicate).   It is also unclear 
exactly what triggers star formation, as the current star formation rate 
does not correlate in any simple way with HI surface density 
(though stars do seem to form where the HI density is highest).  To shed some 
light on these questions, we would like to have a larger sample of galaxies 
to study (so that, for example, we could compare the distribution of 
M(HI)/L$_{\rm B}$ with that seen in giant ellipticals).  Fortunately, the 
recently--completed HI Parkes All--Sky Survey (HIPASS) offers the 
possibility of doing this in the near future. 
\vspace*{-0.1cm}

\section{Finding gas--rich early--type galaxies with HIPASS} 
HIPASS (Barnes et al.\ 2000) is an HI imaging survey of the entire 
southern sky ($\delta < 0^\circ$, but now being extended further north) 
carried out with a 13--beam receiver on the 64\,m Parkes radiotelescope.  
HIPASS spectra cover the velocity range $-1280$ to $+12,700$ km s$^{-1}$.  

A preliminary examination (Sadler 2000) of HIPASS spectra of $\sim2500$ bright
early--type galaxies finds an HI detection rate of at least 5\% for ellipticals
and 12\% for S0 galaxies catalogued in the RC3 (de Vaucouleurs et al.\ 1991).
The detection rate is significantly higher for low--luminosity E and S0 
galaxies than for more luminous ones (22$\pm$5\% for early--type galaxies 
with $-18>$M$_{\rm B}>-20$, versus 10$\pm2$\% for those with 
M$_{\rm B}<-20$), though the RC3 contains relatively few low--luminosity
galaxies.  However, automated galaxy finders will eventually detect many more 
uncatalogued galaxies in the HIPASS data cubes. 

Predictions based on the optical luminosity function
of E/S0 galaxies and the observed HI detection rate (Sadler 1997) suggest that
as many as 100--150 uncatalogued HIPASS detections will be low--luminosity E/SO
galaxies with apparent magnitudes in the range B=15--18. This will provide 
a data set large enough to derive a reliable HI mass function for nearby 
E and S0 galaxies, and to explore in detail the effects of both luminosity 
and environment on the HI content of early--type galaxies. 

\vspace*{-0.1cm}
\section{Conclusions }
Luminous (`boxy') and low--luminosity (`disky') ellipticals appear to differ
not only in their structure and kinematics, but in their HI content and
star--formation history.  HI gas disks are much more common in low--luminosity
ellipticals than in high--luminosity ones, and the HI disks in low--luminosity
ellipticals commonly support a modest level of ongoing massive star formation,
which is not seen in the giant ellipticals. This implies that low--luminosity
ellipticals form much more slowly than giant ellipticals, and in many
cases are still forming stars today.


\vspace*{-0.1cm}
\begin{references} 
\reference
Barnes, D.G. et al.\ 2000, MNRAS, in press 
\reference
Bender, R., D\"obereiner, S., \& M\"ollenhof, C. 1988, \aaps, 74, 385 
\reference 
Binggeli, B., Sandage, A. \& Tammann, G.A. 1988, \araa, 26, 509 
\reference 
Buson, L.M., Sadler, E.M., Zeilinger, W.W., Bertin, G., Bertola, F., Danziger, 
J., Dejonghe, H., Saglia, R.P., \& de Zeeuw, P.T. 1993, A\&A, 280, 409  
\reference 
Capaccioli, M., Caon, N., \& Rampazzo, R. 1990, MNRAS, 24P 
\reference
de Vaucouleurs, G. et al.\ 1991, `Third Reference Catalogue of Bright 
Galaxies', Springer--Verlag  
\reference 
Knapp, G.R., Turner, E.L., \& Cunniffe, P.E. 1985, AJ, 90, 454 
\reference
Kormendy, J.\ \& Bender, R. 1996, ApJ, 464, L119  
\reference 
Lake, G., \& Schommer, R.A. 1984, ApJ, 280, 107 
\reference 
Lake, G., Schommer, R.A., \& van Gorkom, J.H. 1987, ApJ, 314, 57 
\reference
Phillips, M.M., Jenkins, C.R., Dopita, M.A., Sadler, E.M., \& Binette, L. 1986, 
AJ, 91, 1062 
\reference 
Rix, H.-W. \& White, S.D.M. 1990, \apj, 362, 52 
\reference 
Sadler, E.M. \& Gerhard, O.E. 1986, MNRAS, 214, 177 
\reference 
Sadler, E.M. 1997, PASA, 14, 45 
\reference 
Sadler, E.M., Oosterloo, T.A., Morganti, R., \& Karakas, A. 2000, AJ, 119, 1180 
\reference 
Sadler, E.M. 2000, in `Gas and Galaxy Evolution', ed. J.E. Hibbard, M.P. Rupen 
\& J.H. van Gorkom, ASP Conf. Series, in press 
\reference 
Searle, L., Sargent, W.L.W. \& Bagnuolo, W.G. 1973, ApJ, 179, 427
\reference
van den Bergh, S., 1989, \pasp, 101, 1072 
\end{references}
\end{document}